\documentclass{aastex631}
\usepackage{newtxtext,newtxmath}
\usepackage{lastpage}

\usepackage{booktabs}
\usepackage{longtable} % 支持跨页表格
\usepackage{lipsum} % 示例文本
\usepackage{array} % 更多的数组和表格选项
\usepackage{soul} % 支持自动换行的删除线
\usepackage{graphicx}

\begin{document}

\title{Luminosity Function of collapsar Gamma-Ray Bursts:the Progenitor of Long Gamma-Ray Bursts Is Not Singular
}

\correspondingauthor{Yan-Kun Qu}
\email{quyk@qfnu.edu.cn}
\author[0000-0002-9838-4166]{Yan-Kun Qu}
\affiliation{College of Physics and Physical Engineering\\ Qufu Normal University, Qufu,273165, People's Republic of China}
\author{Zhong-Xiao Man}
\affiliation{College of Physics and Physical Engineering\\ Qufu Normal University, Qufu,273165, People's Republic of China}
\author[0000-0003-0672-5646]{Shuang-Xi Yi}
\affiliation{College of Physics and Physical Engineering\\ Qufu Normal University, Qufu,273165, People's Republic of China}
\author[0000-0002-0786-7307]{Yu-Peng Yang}
\affiliation{College of Physics and Physical Engineering\\ Qufu Normal University, Qufu,273165, People's Republic of China}



\begin{abstract}

Gamma-ray bursts (GRBs) are powerful probes of the high-redshift universe. However, the proportion of collapsar GRBs among long GRBs and their event rate relative to the star formation rate (SFR) remain contentious issues. We assume that long GRBs with $z\geq 2$ are all collapsar GRBs and construct the luminosity function using a high-redshift sample from the Swift satellite spanning 2004 to 2019. We model the luminosity function with a broken power-law form and consider three scenarios: no evolution, luminosity evolution, and density evolution. Our results are as follows: 1) The no-evolution model can be ruled out. 2) The fitting results indicate that to adequately explain the observations, a significant redshift evolution in either luminosity (evolution index $\delta = 1.54^{+0.21}_{-0.22}$) or density ($\delta = 2.09^{+0.29}_{-0.26}$) is required.
 This excludes the possibility that the evolution of long GRBs with redshift is due to contamination from non-collapsar GRBs. 3) The luminosity evolution model predicts that the number of collapsar GRBs with $z<2$ and $P \geq 1 \, \text{ph} \, \text{cm}^{-2} \, \text{s}^{-1}$
 is 138.6, accounting for 82.5\% of the observed long GRBs with  $z<2$ and $P \geq 1 \, \text{ph} \, \text{cm}^{-2} \, \text{s}^{-1}$. The density evolution model predicts that  the number of collapsar GRBs with $z<2$ and $P \geq 1 \, \text{ph} \, \text{cm}^{-2} \, \text{s}^{-1}$
 is 80.2, accounting for 47.7\% of the observation. Regardless of the model, a substantial portion of the long GRBs are not collapsar GRBs.

%Whether long gamma-ray bursts (lGRBs) have multiple origins has long been an important and unresolved issue.In this study, we categorize 423 Swift  lGRBs with redshifts from 2004 to 2019 into two groups based on redshift ($> 2$ or not).Among these, 206 have redshifts greater than 2, comprising about 49$\%$ of the total sample. We revisited the luminosity function of lGRBs using the subsample  with redshift greater than 2. Employing the widely accepted broken power-law LF, our study reveals a necessity for significant redshift evolution in luminosity (with an evolution index $\delta$) or density ($\delta$) to effectively explain the observations. This indicates that even after excluding the influence of lGRBs with potentially different origins from low-redshift, there is still notable evolution in long GRBs. By utilizing the best-fit parameters of the luminosity function from high-redshift samples, we calculate the expected number of lGRBs across z from (0-10). The expected number of low-redshift lGRBs is significantly lower than observation, suggesting that only a subset of low-redshift GRBs share the same origin as high-redshift lGRBs, implying that low-redshift lGRBs may have more than one origin.

\end{abstract}

\keywords{Gamma-ray bursts (629) --- Luminosity function(942) --- Star formation (1569)}

\section{Introduction} \label{sec:intro}

  Gamma-ray bursts (GRBs) are extremely luminous astronomical phenomena with a wide redshift distribution, and the highest redshift of GRBs detected so far is $z\sim 9.4$~\citep{2011ApJ...736....7C}. Moreover, it has been predicted that next-generation detectors, such as SVOM, could potentially detect GRBs with a maximum redshift of up to 12 \citep{2024A&A...685A.163L}. Konus-Wind \citep{1981Ap&SS..80....3M} first detected the bimodal distribution of GRBs, long and short bursts based on their $T_{90}$ duration, which is then confirmed by BATSE~\citep{1993ApJ...413L.101K} and Swift \citep{2013ApJ...764..179B}. As a result, GRBs are commonly divided into long and short bursts based on their $T_{90}$ duration. Long GRBs are believed to originate from the collapse of massive stars, as they often occur in star-forming regions \citep{1997ApJ...486L..71T,1998ApJ...507L..25B}, associating with Type $I_{c}$ supernovae \citep{1998Natur.395..670G,2008ApJ...687.1201K,2010MNRAS.405...57S}, and have a strong correlation with bright ultraviolet regions within their host galaxies \citep{2004A&A...425..913C}. Short GRBs, on the other hand, are thought to arise from binary mergers, as they often occur in regions far from star-forming areas \citep{2014ARA&A..52...43B}, associated with gravitational waves~\citep{2017ApJ...848L..13A} and kilonovae \citep{2013Natur.500..547T,2019MNRAS.489.2104T,2020NatAs...4...77J}. However, many exceptions have been reported. For example, GRB 211211A \citep{2022Natur.612..223R}, with  $T_{90} > 30s$ and redshift $z = 0.076$, should be a typical long burst, but it was found associated with a kilonova. \cite{2022Natur.612..232Y} suggested that the progenitor of GRB 211211A is likely a white dwarf-neutron star merger system. GRB 230307A \citep{2024Natur.626..737L}, with $z = 0.0646$ and $T_{90} \sim 35s$, was identified from a binary neutron star merger. \cite{2013ApJ...764..179B} found that a considerable proportion of short GRBs should originate from the collapse of massive stars.

Long GRBs are believed to be produced by the supernova (SN) explosions of massive Wolf–Rayet stars, with their event rate expected to parallel the star formation rate (SFR). However, there is considerable debate on this issue. A few studies found no discrepancy between GRB event rate and SFR~\citep{2012A&A...539A.113E,2013ApJ...772...42H,2016A&A...587A..40P}. Some researches indicate that the GRB event rate significantly exceed the expectation at high redshifts~\citep{2008ApJ...673L.119K,2008ApJ...683L...5Y,2009MNRAS.400L..10W,2011MNRAS.417.3025V}, while others found an excess at low redshifts ~\citep{2015ApJ...806...44P,2015ApJS..218...13Y,2017ApJ...850..161T,2019MNRAS.488.5823L,2022MNRAS.513.1078D}.

Many theories have been proposed to explain the significant deviation between the long GRBs event rate and the SFR, including the evolution of the initial mass function of stars~\citep{2008AcASn..49..387X,2011ApJ...727L..34W,2022ApJ...938..129L}, the evolution of cosmic metallicity~\citep{2008MNRAS.388.1487L,2010MNRAS.406..558Q} , and superconducting cosmic strings \citep{2010PhRvL.104x1102C,2021MNRAS.508...52L} found that the luminosity or density evolution, or even a unusual ``triple power-law" luminosity function, is necessary to explain the observations of long GRBs.~\cite{2023ApJ...958...37D} categorized GRBs by luminosity and found that high-luminosity GRBs rate follows the SFR, while the event rate of low-luminosity GRBs significantly exceeds the SFR at low redshifts. They proposed that long bursts should have two different progenitor types. \cite{2024ApJ...963L..12P} found that if the long GRB event rate is subtracted from the SFR, the remaining portion closely resembles the event rate of short GRBs. Based on this, they suggested that for the redshifts $z \geq 2$, all long GRBs originate from the collapse of massive stars, whereas for the redshifts $z<2$, a substantial portion, approximately $60\%$, may originate from binary neutron star mergers.

If long GRBs at $z \geq 2$ are purely from the collapse of massive stars (hereafter abbreviated as collapsar GRBs), the luminosity function of collapsar GRBs can be constructed using the high-redshift long GRB samples. In this Letter, we construct the luminosity function of collapsar GRBs, study its evolution, and investigate the proportion of collapsar GRBs among all long GRBs. In Chapter 2, we describe the samples and methods used in this work. We then present the luminosity function fitting results in chapter 3. In Chapter 4, a brief summary and discussion are presented. In this paper, we adopt a flat $\Lambda$CDM cosmological model with $H_0 = 70 \, \text{km} \, \text{s}^{-1} \, \text{Mpc}^{-1}$, $\Omega_m = 0.3$, and $\Omega_\Lambda = 0.7$.

\section{Samples and Methods } \label{sec:style}
\subsection{Samples  }
Since 2004, when the Swift satellite was launched, nearly 2000 GRBs have been detected, and most of which are long GRBs with a smaller portion being short GRBs. Approximately, one-third of the Swift GRBs have redshift measurements, and it is significantly higher than the redshift detection rate of previous satellites like BATSE. This provides a solid foundation for studying the luminosity function of GRBs.
One complex issue that researchers face when studying the luminosity function of GRBs is instrumental selection effect. The Swift/BAT trigger system is intricate, and its sensitivity to GRBs cannot be precisely parameterized \citep{2006ApJ...644..378B}. In practice, not all less power GRBs with peak fluxes slightly above the instrument threshold can be successfully triggered.\cite{2021MNRAS.508...52L} assumed that the intrinsic peak flux distribution of GRBs follows a broken power law . Their findings indicate that the observed peak flux distribution of long GRBs with peak flux \( P > 1 \, \text{ph} \, \text{cm}^{-2} \, \text{s}^{-1} \) aligns well with the broken power law, whereas those with \( P < 1 \, \text{ph} \, \text{cm}^{-2} \, \text{s}^{-1} \) show a clear deviation from this model. This suggests that above this flux limit, the instrumental selection effects influencing the incomplete sampling are negligible.

There is still considerable controversy regarding the relationship between the long GRB event rate and the SFR. \cite{2024ApJ...963L..12P} found that long GRBs with $z\geq 2$ better match the SFR, suggesting that a significant portion of these high-redshift long GRBs originate from stellar collapse. Based on this view, we hypothesize that long GRBs with $z\geq 2$ could originate from massive stellar collapses, and supposed as collapsar GRBs. In this work, for distinction, we refer to GRBs with $T_{90} \geq 2s$ as long GRBs.

\begin{figure}[ht!]
\plotone{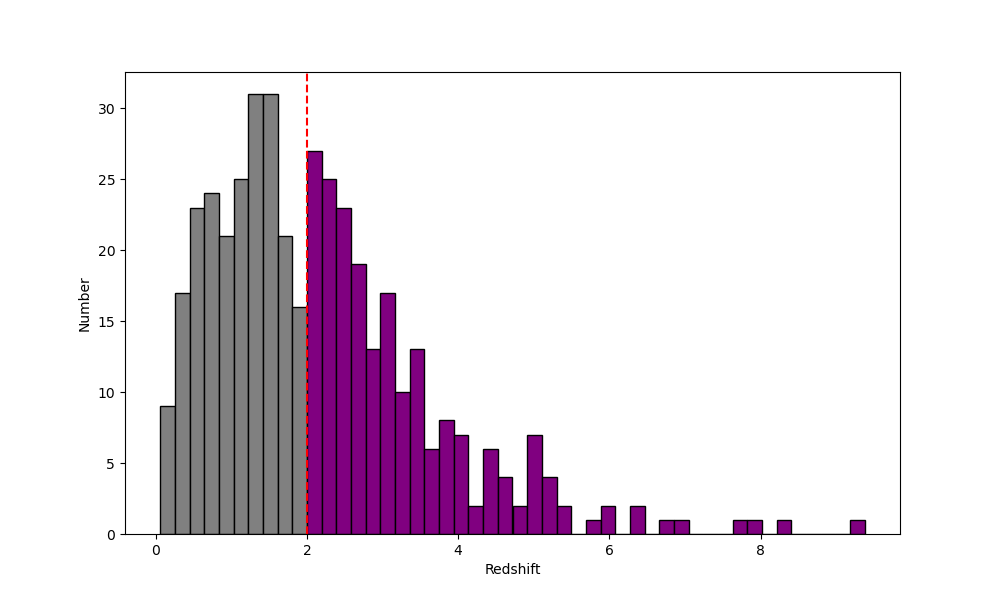}
\caption{The distribution of redshifts for all 423 Swift long GRBs, with the red dashed line representing the location of $z=2$, and there are a total of 206 samples with  $z\geq 2$.
\label{fig:z}}
\end{figure}

\begin{figure}[ht!]
\plotone{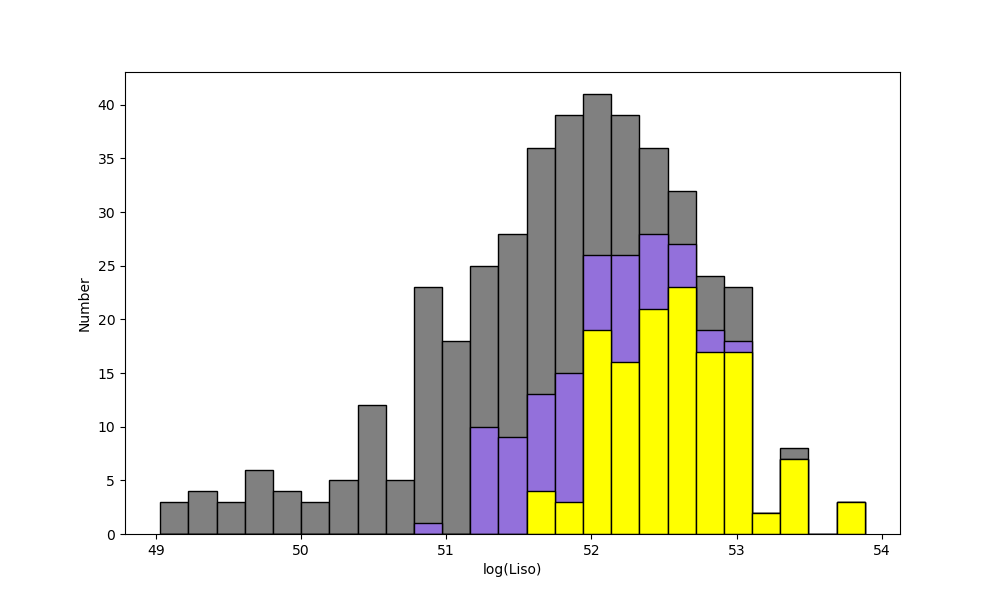}
\caption{The $L_{iso}$ distribution of GRBs, where gray represents  all 423 Swift long GRBs , purple represents 206 bursts with $z\geq 2$, and yellow represents bursts with $z\geq 2$ and 1-s peak flux $P \geq 1$ photons cm$^{-2}$ s$^{-1}$ , totaling 134 bursts. 
\label{fig:Liso}}
\end{figure}

As of November 2019, excluding low-luminosity GRBs with \( L < 10^{49} \) erg s\(^{-1}\), there were 423 Swift long GRBs with redshift measurements.\footnote{It should be noted that compared to \cite{2021MNRAS.508...52L}, the GRB100724A, which has $T_{90}= 1.39s$ and should be classified as a short burst, is not included in our sample.}
 Among these, 301 had peak fluxes $P \geq 1 \, \text{ph} \, \text{cm}^{-2} \, \text{s}^{-1}$ , resulting in a redshift completeness of $42\%$. Using this dataset, \cite{2021MNRAS.508...52L} found that either luminosity evolution or density evolution is required for the luminosity function to  fit the sample, and a triple power-law model outperforms a broken power-law model. To directly compare with \cite{2021MNRAS.508...52L}'s results, we also used the GRB samples detected up to November 2019 with peak fluxes  $P \geq 1 \, \text{ph} \, \text{cm}^{-2} \, \text{s}^{-1}$ . There are totally 206 GRBs with $z\geq 2$, accounting for $49\%$ of the known redshift GRBs, of which 134 have peak fluxes  $P \geq 1 \, \text{ph} \, \text{cm}^{-2} \, \text{s}^{-1}$ . The distribution of redshift $z$ and luminosity $L_{\rm iso}$ can been seen in Figs. \ref{fig:z} and \ref{fig:Liso}.

\subsection{ANALYSIS METHOD}

\begin{figure}[ht!]
\plotone{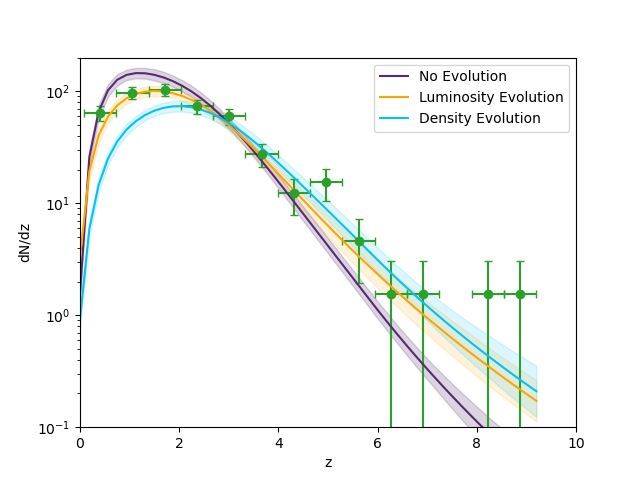}
\caption{
 The redshift distribution. Three curves representing the redshift distributions of collapsar GRBs calculated by three different models. The shaded areas indicate the 1-sigma confidence intervals. The green data points represent the observed redshift distribution of 301 long GRBs with $P \geq 1$ photons cm$^{-2}$ s$^{-1}$.
 }
\label{fig:dNdz}
\end{figure}

\begin{figure}[ht!]
\plotone{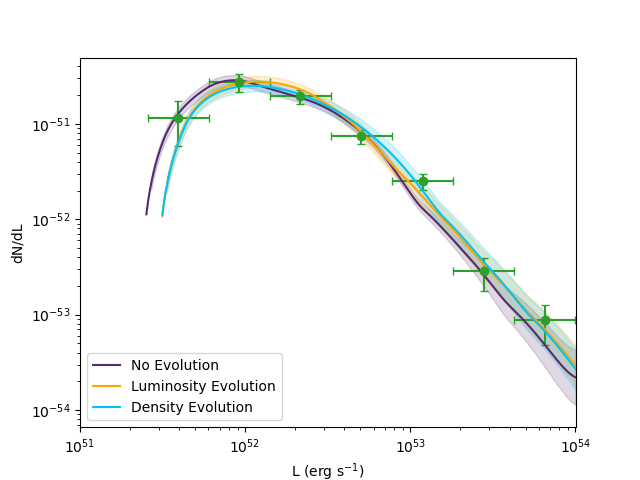}
\caption{The luminosity distribution of 134 long GRBs with $z \geq 2$ and \( P \geq 1 \) photons cm\(^2\) s\(^{-1}\). The three curves represent the best-fit expected luminosity distributions of three different models. The shaded areas indicate the 1-sigma errors.
 }
\label{fig:dNdL_zGe2}
\end{figure}

\begin{figure}[ht!]
\plotone{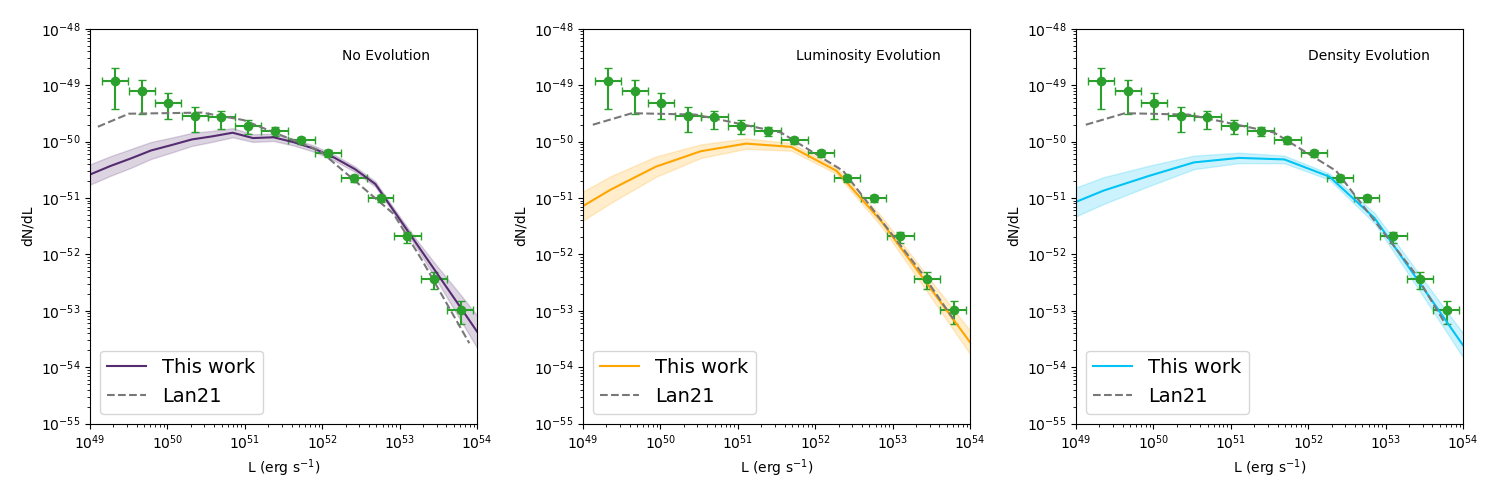}
\caption{The green data points represent the luminosity distribution of 301 long GRBs with $P \geq 1 \ \text{photon cm}^{-2} \ \text{s}^{-1}$. The solid lines depict the luminosity distribution of collapsar GRBs as predicted by different models. For comparison, we also plot the luminosity distribution of long GRBs predicted by different models from \cite{2021MNRAS.508...52L}  (gray dashed lines). From left to right, the models are: No Evolution Model, Luminosity Evolution Model, and Density Evolution Model.
 }
\label{fig:dNdL}
\end{figure}

The maximum likelihood method (MLE) was first introduced by Ronald A. Fisher in the early 20th century ~\citep{fisher1921probable,fisher1925statistical}.We adopted a maximum likelihood method  introduced by ~\cite{1983ApJ26935M} 
It is an effective method for optimizing the free parameters of  the luminosity function, as demonstrated by numerous previous literature \citep{2019MNRAS.490..758Q,2006ApJ...643...81N,2009ApJ...699..603A,2012ApJ...751..108A,2010ApJ...720..435A,2014MNRAS.441.1760Z,2016MNRAS.462.3094Z,2019MNRAS.488.4607L,2021MNRAS.508...52L,2022ApJ...938..129L}.
For our purposes, the maximum likelihood can be written as follows:

\begin{equation}
    \mathcal{L}=e^{(-N_{exp})} \prod_{i=1}^{N_{obs}} \Phi(L_{i},z_{i},t_{i}) 
    \label{eq:L}
\end{equation} 
where $N_{exp} $ represents the expected detection number of GRBs, $N_{obs}$ is the observed sample size, and $ \Phi(L, z, t) $ denotes the observed rate of GRBs per unit time, with redshift in the range  $z$  to  $z + dz$ , and luminosity  in the range  L  to  L + dL . The specific form of  $\Phi(L, z, t) $ is:

\begin{equation}
\begin{split}
    \Phi(L, z, t) &= \frac{d^{3} N}{dtdzdL} \\
    &= \frac{d^{3}N}{dtdVdL} \times \frac{dV}{dz} \\
    &= \frac{\Delta \Omega}{4\pi}  \theta(P)  \frac{\psi(z)}{(1+z)}  \phi(L,z) \times \frac{dV}{dz}
\end{split}
\label{eq:Phi}
\end{equation}
Where $\Delta \Omega = 1.33 \, \text{sr}$ indicates the field of view of the Swift/BAT instrument. The term $\theta(P) \equiv \theta_{\gamma}(P) \theta_{z}(P)$ represents the detection efficiency, which is the probability of detecting a burst and measuring its redshift for a given peak flux $P$. Our sample is selected based on  $P \geq 1 \, \text{ph} \, \text{cm}^{-2} \, \text{s}^{-1}$ , so we assume $\theta_{\gamma}(P)=1$. For $\theta_{z}(P)$, we utilize the results from \cite{2021MNRAS.508...52L} , specifically $\theta_{z}(P)=1/(1+(2.09 \pm 0.26) \times (0.96 \pm 0.01)^P)$.
The function $\psi(z)$ denotes the comoving event rate of GRBs. The factor $(1 + z)^{-1}$ accounts for cosmological time dilation. The function $\phi(L,z)$ is the normalized GRB luminosity function (LF), which may evolve with redshift depending on the model  (details provided below). The term $ {dV(z)}/{dz} = {4\pi c D_{L}^{2}(z)}/{H_{0}(1 + z)^2 \sqrt{\Omega_{m}(1 + z)^3 + \Omega_{\Lambda}}}$ represents the comoving volume element in a flat $\Lambda$CDM model, where $D_{L}(z)$ is the luminosity distance at redshift $z$.

According to the collapsar origin hypothesis, the occurrence of each GRB indicates the death of a short-lived massive star. Therefore, the GRB formation rate, $\psi(z)$, should be intrinsically related to the cosmic SFR, $\psi_{\star}(z)$, such that $\psi(z) = \eta \psi_{\star}(z)$. Here, $\eta$ denotes the GRB formation efficiency. The SFR, $\psi_{\star}(z)$(in units of $ M_{\odot} \, \text{yr}^{-1} \, \text{Mpc}^{-3}$), can be written as \citep{2006ApJ...651..142H,2008MNRAS.388.1487L},

\begin{equation}
 \psi_{\star}(z)=\frac{0.0157+0.118z}{1+(z/3.23)^{4.66}}
\end{equation}

For the GRB luminosity function $\phi(L, z)$, we utilize a general expression in form of  broken power-law:

\begin{equation}
\phi(L,z)=\frac{A}{\ln(10)L}\left\{
\begin{array}{ll}
\left(\frac{L}{L_{c}(z)}\right)^{a}, & L \leq L_{c}(z) \\
\\
\left(\frac{L}{L_{c}(z)}\right)^{b}, & L > L_{c}(z)
\end{array}
\right.
\end{equation}
where $A$ is a normalization constant, $a$ and $b$ are the power-law indices for luminosities below and above the break luminosity $L_{c}(z)$, respectively.

Taking into account the flux threshold set for a $100\%$ trigger efficiency (i.e., $P_{\text{lim}} = 1 \, \text{ph} \, \text{cm}^{-2} \, \text{s}^{-1}$ within the $15-150 \, \text{keV}$ energy range), the expected number of GRBs can be described as

%Taking into account the flux threshold established for a $100\%$ trigger efficiency (i.e., $P_{\text{lim}} = 1 \, \text{photons} \, \text{cm}^{-2} \, \text{s}^{-1}$ within the $15-150 \, \text{keV}$ energy range), the anticipated number of GRBs can be described as 

\begin{equation}
\begin{split}
    N_{exp}=\frac{\Delta \Omega T}{4 \pi}\int_{z_{min}}^{z_{max}}\int_{max[L_{min},L_{lim}(z)]}^{L_{max}}\theta(P(L,z))\frac{\psi(z)}{1+z}\\ \times \phi(L,z) dLdV(z)
\end{split}
\label{eq:Nexp}
\end{equation}

In our study, Swift's mission duration is approximately $T \sim 15$ years, covering our sample period. Given that the redshift for the current BAT sample is $z < 10$, the maximum redshift analyzed is $z_{max}=10$, and we set the minimum redshift as $z_{\min} = 2$.
We assume the luminosity function extends from $L_{min}=10^{49}\text{ erg} \, \text{s}^{-1}$ to $L_{max}=10^{55}\text{ erg} \, \text{s}^{-1}$ \citep{2015MNRAS.447.1911P}. The luminosity threshold mentioned in Equation \ref{eq:Nexp} can be estimated as follows:

\begin{align}
   L_{lim}(z)=4\pi D_{L}^{2}(z)P_{lim}\frac{\int_{1/(1+z) \text{ keV}}^{10^{4}/(1+z)\text{ keV}}E N(E)dE}{\int_{15 \text{ keV}}^{150 \text{ keV}}N(E)dE}
\end{align}

The GRB photon spectrum, \( N(E) \), is typically modeled using the Band function \citep{1993ApJ...413..281B,2006ApJS..166..298K}, characterized by low-energy and high-energy spectral indices of \(-1\) and \(-2.3\), respectively.

Considering (i) the potential bias in the connection between the GRB formation rate and the cosmic SFR, and (ii) the GRB luminosity function (LF) may evolve with redshift, we introduce an additional evolution factor $(1 + z)^\delta$ as shown in the aforementioned equation, where $\delta$ is a free parameter. In this work, we consider three different models, (i) the GRB formation rate strictly follows the SFR, i.e., $\psi(z) = \eta\psi (z)$, and the LF does not evolve with redshift, i.e., $L_{ci} (z) = L_{ci,0} = \text{constant}$. (ii) while the GRB formation rate still scales with the SFR, the break luminosity in the GRB LF increases with redshift, i.e., $L_{ci} (z) = L_{ci,0}(1 + z)^\delta$. (iii) the formation rate follows the SFR and incorporates an additional evolution factor, i.e., $\psi(z) = \eta\psi (z)(1 + z)^\delta$, while the LF is considered unevolved. 

For a given model, the free parameters can be optimized by maximizing the likelihood function (Equation \ref{eq:L}) Given the multiple parameters in our models, we employ Markov Chain Monte Carlo (MCMC) sampling techniques, e.g., the public code EMCEE \citep{2013PASP..125..306F}, to derive the best fit values and uncertainties of model parameters.

After determining the optimal parameters , we can estimate the predicted number of collapsar GRBs within the redshift range of 0 to 2 for each model  by substituting these   parameters into Equation \ref{eq:Nexp} and setting the integration limits for redshift to 
$z_{min}=0$ and $z_{max}=2$

\section{RESULTS} \label{sec:RESULTS}

Using the MCMC method, we obtained the best-fitting parameters for three models, and the results are shown in table \ref{table1}. We also present the log-likelihood values and the AIC for comparing different models. AIC is defined as $AIC = -2 \ln \mathcal{L} + 2n$, where $\ln \mathcal{L}$ is the log-likelihood value, and $n$ is the number of free parameters in the model. After obtaining the AIC for each model, we can use the Akaike weight $exp(-AIC_i/2)$ to indicate the confidence of one model relative to another, defined specifically as follows,
\begin{equation}
\begin{split}
    P(M_{i})=\frac{e^{-AIC_{i}/2}}{e^{-AIC_{1}/2} + e^{-AIC_{2}/2}}
\end{split}
\label{eq:aIC}
\end{equation}

In Fig \ref{fig:dNdL_zGe2}, we plot the luminosity distribution of 134 GRBs with $z\geq 2$ and $P \geq 1 \, \text{ph} \, \text{cm}^{-2} \, \text{s}^{-1}$ and in Fig \ref{fig:dNdL}, we plot the luminosity distribution of all 301 GRBs with  $P \geq 1 \, \text{ph} \, \text{cm}^{-2} \, \text{s}^{-1}$ and the luminosity distribution of collapsar GRBs as predicted by different models.For comparison, we also plot the luminosity distribution of long GRBs predicted by different models from \cite{2021MNRAS.508...52L}.

No evolution model implies that the luminosity function does not evolve with redshift, and the  GRB events rate strictly traces the SFR. Fig \ref{fig:dNdz} shows the number of low-redshift collapsar GRBs predicted by the no evolution model is significantly higher than the observed number of long GRBs, which is unacceptable. Interestingly, this result is quite similar to the findings by \cite{2021MNRAS.508...52L} using all GRBs with $P \geq 1 \, \text{ph} \, \text{cm}^{-2} \, \text{s}^{-1}$ . Using equation \ref{eq:aIC}, the  probability of the no evolution model being correct relative  to the luminosity evolution  model is only $10^{-3}$. 

We also computed the BIC values\citep{24ce203a-855a-3aa9-952f-976d23b28943} and included them in Table \ref{table1} . The results indicate that the difference in BIC (\(\Delta \text{BIC}\)) between the no evolution model and the luminosity evolution model is 10.7, and the difference between the no evolution model and the density evolution model is 12.04. Based on these results, the no evolution model can be confidently ruled out, as the established threshold for model comparison (\(\Delta \text{BIC} > 8\)) is exceeded.
Therefore, we suggest that the no evolution model is insufficient for fitting the GRB data.

The luminosity evolution model implies that the GRB events rate follows the SFR, but the break luminosity of GRBs evolves with redshift, $L_c(z) = L_{c,0} (1 + z)^\delta$. 
Our results indicate a strong evolution parameter with $\delta = 1.54^{+0.21}_{-0.22}$, which is consistent with previous findings $\delta = 0.5 \sim 2$ and indicates that GRBs at higher redshifts are brighter than those at lower redshifts.  Using the best-fitting parameters of the luminosity evolution model, we calculate the expected number of collapsar GRBs within the redshift range (0, 2). With the integral limits of equ \ref{eq:Nexp} adjusted to (0, 2) while keeping the best-fitting parameters unchanged, the luminosity evolution model predicts that the number of collapsar GRBs with $z<2$ and $P \geq 1 \, \text{ph} \, \text{cm}^{-2} \, \text{s}^{-1}$
 is 138.6, accounting for 82.5\% of the observed long GRBs. This suggests that $17.5\%$ of low-redshift long GRBs should not be collapsars.

The density evolution model represents that the luminosity function does not evolve with redshift, but the ratio of GRB event rate to SFR is not constant instead  of changing with redshift, $\psi(z) = \eta \psi^*(z)(1 + z)^\delta$. We found a non zero value of $\delta = 2.09^{+0.29}_{-0.26}$, indicating that the observations data can also be explained by the density evolution model. Using the best-fitting parameters of the Density evolution model, we calculate the expected data of collapsar GRBs within the redshift range (0, 2). With the integral limits of equation \ref{eq:Nexp} adjusted to (0, 2) while keeping the best-fitting parameters unchanged, the density evolution model predicts that  the number of collapsar GRBs with $z<2$ and $P \geq 1 \, \text{ph} \, \text{cm}^{-2} \, \text{s}^{-1}$
 is 80.2, accounting for 47.7\% of the observation.This suggests that $52.3\%$ of low-redshift long GRBs may not be collapsars, which is consistent with the reults given by \cite{2024ApJ...963L..12P}, where they found that approximately $60\%$ of long GRBs with $z<2$ are not collapsar GRBs.

\section{Conclusion and Discussion} \label{sec:cite}
In this work, we discovered significant evolution with redshift in the sample of collapsar GRBs. Specifically, both the luminosity evolution model and the density evolution model significantly outperform the no evolution model. Previous research suggested that the observed evolution of long GRBs with redshift might be due to the inclusion of non-collapsar GRBs in the sample. Our study challenges this issue by showing that even after excluding non-collapsar GRBs and using a pure sample of collapsar GRBs, significant evolution is still present. This indicates that the evolution of GRBs with redshift is an inherent property. However, it remains to be determined whether this evolution is intrinsic to GRBs themselves or a result of cosmological effects.  Future research could investigate the redshift evolution of other high-redshift celestial sources, such as BL Lac objects and FSRQs, to compare the similarities and differences in the evolution of various types of high-energy sources and draw more comprehensive conclusions

Regardless of using luminosity or density evolution models, the predicted number of low-redshift collapsar GRBs is far smaller than the observed number of low-redshift long GRBs.
This provides a strong evidence that a significant portion of low-redshift long GRBs does not originate from massive star collapses.
It is noteworthy that the predicted number of non-collapsar long GRBs differs significantly between the two evolution models. The density evolution model predicts that $52.3\%$ of long GRBs with $z<2$ are non-collapsar GRBs, which is consistent with the conclusions of \cite{2024ApJ...963L..12P}. In contrast, the luminosity evolution model predicts that $17.5\%$ are non-collapsar GRBs. A vexing issue is that the goodness of fit for both the luminosity and density evolution models is quite similar, making it impossible to distinguish and exclude one model over the other. This has been a longstanding problem, with many previous studies finding similar results. However, when determining the proportion of low-redshift non-collapsar long GRBs among long GRBs, distinguishing between these two models becomes crucial. Future research that successfully differentiates these models will have significant implications for studying the proportion of low-redshift non-collapsar long GRBs.

Notably, 
\cite{2018ApJ...852....1Z} proposed that when short GRBs are classified by luminosity, there are significant differences in the luminosity function and event rate between high-luminosity and low-luminosity GRBs. \cite{2023ApJ...958...37D} applied a similar approach to long GRBs and  categorized long GRBs into high-luminosity and low-luminosity types, finding that the event rate of high-luminosity long GRBs closely follows the SFR, while the event rate of low-luminosity long GRBs significantly deviates from the SFR. Our study can well explain \cite{2023ApJ...958...37D}'s findings. The reason of high-luminosity GRBs closely following the SFR is that the main component of high-luminosity GRBs is long GRBs with $z \geq 2$, which are collaspar GRBs, thus closely following the SFR. Meanwhile, many non-collaspar GRBs are mixed into low-luminosity GRBs, resulting in a significant deviation from the SFR. To further verify this conclusion, long GRBs with $z\geq 2$ can be classified by high and low luminosity, and their event rates can be calculated separately to verify their relationship with the SFR. We predict that for GRBs with $z\geq 2$, both high-luminosity and low-luminosity GRBs will follow the SFR, and there will no longer be a significant deviation of low-luminosity GRBs from the SFR due to the sample with $z\geq 2$ consisting entirely of collaspar GRBs.

In previous works, such as \cite{2021MNRAS.508...52L}, the authors used a triple power-law for relevant calculations. As shown in the right panel of Figure 3 in \cite{2021MNRAS.508...52L}, the luminosity function constructed using a broken power-law, whether with or without evolution terms, significantly deviates from the observed values at low luminosities in the $L_{iso}$ distribution. Generally, triple power-law is considered less physical, and luminosity functions are typically constructed using power-law, broken power-law, or power-law with a cutoff. However, \cite{2021MNRAS.508...52L} found that the luminosity function constructed using triple power-law significantly outperformed broken power-law. Our study can provide a reasonable explanation for this issue: long GRBs consist of two components. One component is collaspar GRBs, whose luminosity function follows a broken power-law with overall higher luminosity. Another component is non-collaspar GRBs, whose luminosity function also follows a broken power-law but with overall lower luminosity. When these two types of GRBs are mixed together, they form the observed long GRBs, making it appear that a triple power-law is needed to better fit the luminosity function of long GRBs.

The progenitor stars of non-collaspar GRBs within long GRBs remain a puzzle that has perplexed researchers for years. It has been proposed that the components of long GRBs might not be singular. \cite{2023ApJ...958...37D} suggested that this component might originate from the merger of binary compact stars. \cite{2024ApJ...963L..12P} subtracted the SFR from the event rate of long GRBs, discovering that the residual part closely resemble the merger of binary compact stars, which they regarded as strong evidence that the progenitor of non-collaspar GRBs within long GRBs are binary compact star mergers. 
However, as shown in this work, pure collaspar GRBs do not strictly follow the SFR but exhibit some evolutionary feature. Therefore, directly using the long GRB event rate minus the SFR to obtain the event rate of non-collaspar GRBs is not very rigorous. After constructing the luminosity function of collaspar GRBs, we can easily derive the $L_{iso}$ and $z$ distribution maps of non-collaspar GRBs. Then we can constrain the luminosity function of non-collaspar GRBs. Comparing the luminosity function of non-collaspar GRBs among long GRBs with that of short GRBs will be strong evidence in determining whether the non-collaspar GRBs are indeed binary compact star merger GRBs. If non-collaspar GRBs are confirmed to be binary compact star merger GRBs, the number of binary compact star merger GRBs will be several times larger than the currently recognized. This will have a significant impact on using GRBs to estimate gravitational wave event rate. 

The SVOM satellite has been launched this year (2024), and it is expected that GRBs with $z\ge 12$ can be observed. Moreover, high-quality observational data will significantly advance GRB cosmology, and we look forward to new data validating our conclusions.

\renewcommand{\arraystretch}{1.5}
\begin{table}[ht]
\centering
\caption{}
\begin{tabular}{@{}cccccccccc@{}}
\toprule
Model & Evolution parameter & $\eta$ & $a$ & $b$  & log $L_c$  & ln $\mathcal{L}$ & AIC& BIC \\ 
 &  &  ($10^{-8} M^{-1} _{\odot}$) &  & & (erg s$^{-1}$)&   & & \\ \midrule

No evolution &… & $8.47^{+0.98}_{-0.97}$ & $-0.17^{+0.05}_{-0.05}$ & $-1.67^{+0.24}_{-0.24}$  & $52.92^{+0.11}_{-0.11}$ & -58.46 & 124.92 &136.51\\
Luminosity evolution & $\delta = 1.54^{+0.21}_{-0.22}$ & $5.57^{+0.67}_{-0.69}$ & $-0.13^{+0.06}_{-0.06}$ & $-1.20^{+0.17}_{-0.16}$ & $51.58^{+0.19}_{-0.20}$ & -50.66 & 111.32&125.81 \\
Density evolution & $\delta = 2.09^{+0.29}_{-0.26}$ & $6.81^{+0.72}_{-0.65}$ & $-0.38^{+0.04}_{-0.04}$ & $-1.65^{+0.25}_{-0.24}$ & $52.92^{+0.09}_{-0.11}$ & -49.99 & 110.18&124.47 \\ 

\bottomrule
\end{tabular}
\begin{minipage}{\textwidth}
\footnotesize
\textbf{Note.} The parameter values were determined as the medians of the best-fitting parameters from the Monte Carlo sample. The errors represent the 68\% containment regions around the median values.
\end{minipage}
\label{table1}
\end{table}

%% Also note that the akcnowlodgment environment does not support long amounts of text. If you have a lot of people and institutions to acknowledge, do not use this command. Instead, create a new \section{Acknowledgments}.
\begin{acknowledgments}
We sincerely thank Lan, GX for the valuable discussions and suggestions during the preparation of this manuscript. Additionally, we acknowledge the support from the Shandong Provincial Natural Science Foundation (Grant No. ZR2021MA021).

\end{acknowledgments}


\vspace{5mm}
\facilities{ Swift(XRT and UVOT)}


\software{astropy\citep{2013A&A...558A..33A,2018AJ....156..123A,astropy:2022}}, Emcee\citep{2013PASP..125..306F} 



\begin{table}[htbp]
\centering
\rotatebox{90}{%
\parbox{\textheight}{%
\caption{Example of GRB parameters in online only tables}
\centering
\begin{tabular}{l c c c c c c c c c c c c c c}
\toprule
Name    & Peak Flux & Peak Flux Error & Redshift & $\alpha$ & $\alpha_{err up}$ & $\alpha_{err down}$ & $\beta$ & $\beta_{err up}$ & $\beta_{err down}$ & $E_p$ & $E_p$ err up  & $E_p$ err down & $\log L$ & $\log L$ err \\ 
        & (photons cm$^{-2}$ s$^{-1}$) &  &  &  &  &  &  &  &  & (keV) &  &  & (erg s$^{-1}$) &  \\ \midrule
041228   & 1.61      & 0.25            & 2.30     & -1.54    & 0.05          & -0.05           & -99     & -99          & -99            & 240.00   & 381.06         & -66.53         & 52.09    & 0.07           \\ \midrule
050319   & 1.52      & 0.21            & 3.24     & -2.00    & 0.19          & -0.21           & -99     & -99          & -99            & 44.73 & 27.49          & -43.11         & 52.54    & 0.06           \\ \midrule
050401   & 10.70     & 0.92            & 2.90     & -1.44    & 0.08          & -0.08           & -99     & -99          & -99            & 165.49 & 391.26         & -48.08         & 53.09    & 0.04           \\ \midrule
050502B  & 1.42      & 0.13            & 5.20     & -1.60    & 0.06          & -0.06           & -99     & -99          & -99            & 100.00   & 228.03         & -19.96         & 52.78    & 0.04           \\ \midrule
050505   & 1.85      & 0.31            & 4.27     & -1.41    & 0.12          & -0.12           & -99     & -99          & -99            & 140.25 & 343.09         & -42.77         & 52.70    & 0.07           \\ \midrule
050603   & 21.50     & 1.07            & 2.821    & -1.03    & 0.11          & -0.11           & -2.03   & 0.17         & -0.29          & 343.70  & 87.00            & -87.00            & 53.73    & 0.02           \\ \midrule
050820A  & 2.45      & 0.23            & 2.612    & -1.20    & 0.06          & -0.06           & -99     & -99          & -99            & 490.00    & 435.50          & -181.46        & 52.64    & 0.04           \\ \midrule
050922B  & 1.01      & 0.36            & 4.50     & -1.10    & -99           & -99             & -2.10   & 0.18         & -0.18          & 38.00     & 15.73          & -22.38         & 52.49    & 0.15           \\ \midrule
050922C  & 7.26      & 0.32            & 2.198    & -0.83    & 0.23          & -0.26           & -99     & -99          & -99            & 130.00    & 51.00           & -27.00          & 52.56    & 0.02           \\ \midrule
051008   & 5.44      & 0.35            & 2.77     & -0.80    & 0.23          & -0.22           & -99     & -99          & -99            & 234.76 & 315.18         & -70.12         & 52.89    & 0.03           \\ \bottomrule
\end{tabular}
\parbox{\textwidth}{ \raggedright \textbf{Note:} A value of -99 indicates that this parameter is missing.}

}
}
\end{table}


\bibliography{sample631}{}
\bibliographystyle{aasjournal}



\end{document}